\documentclass[conference, 10pt]{IEEEtran}

\usepackage{multicol, latexsym, amsmath, amssymb}
\usepackage{amssymb}
\usepackage{amsmath}
\usepackage{cite}
\usepackage{algorithm}
\usepackage{algorithmic}
\usepackage{romannum}
\usepackage[dvips]{graphicx}
\usepackage{stfloats}
\usepackage{comment}
\usepackage{latexsym,epsf,times,amsmath,color,amssymb,indentfirst,subfigure,fancyhdr,colortbl,bm,caption}
\usepackage{xspace}
\usepackage{graphicx}
\usepackage{bbm}
\usepackage{graphicx}
\usepackage[dvipsnames]{xcolor}

\setbox1=\hbox{$\bullet$}
\usepackage{stfloats}
\ifCLASSINFOpdf

\else

\fi

\makeatletter
\newcommand*{\rom}[1]{\expandafter\@slowromancap\romannumeral #1@}
\makeatother

\begin{document}

\pagenumbering{arabic}
%
\title{Clustering and Power Optimization for NOMA Multi-Objective Problems}

\author{\IEEEauthorblockN{ Zijian Wang\dag, Mylene Pischella{*}, Luc Vandendorpe\dag}
\IEEEauthorblockA{\dag Institute of Information and Communication Technologies, Electronics and Applied Mathematics\\Universit\'{e} catholique de Louvain,
	Louvain-la-Neuve, Belgium,\\ {*}  CEDRIC, CNAM, Paris, France\\Emails: \{zijian.wang, luc.vandendorpe\}@uclouvain.be, mylene.pischella@cnam.fr
}
}

\maketitle

\newtheorem{theorem}{Theorem}
\newtheorem{lemma}{Lemma}
\newtheorem{proposition}{Proposition}
\newtheorem{corollary}{Corollary}

\begin{abstract}
This paper considers  uplink multiple access (MA) transmissions, where the MA technique is  adaptively selected between Non Orthogonal Multiple Access (NOMA) and Orthogonal Multiple Access (OMA). Two types of users, namely Internet of Things (IoT) and enhanced mobile broadband (eMBB) coexist with different metrics to be optimized, energy efficiency  (EE) for IoT and spectral efficiency (SE) for eMBB. The corresponding multi-objective power allocation problems aiming at maximizing a weighted sum of EE and SE are solved for both NOMA and OMA. Based on the identification of the best MA strategy, a clustering algorithm is then proposed to maximize the multi-objective metric per cluster as well as NOMA use. The proposed clustering, power allocation and MA selection algorithm is shown to outperform other clustering solutions and non-adaptive MA techniques. 
\end{abstract}

\IEEEpeerreviewmaketitle
\section{Introduction}
Power Domain Non Orthogonal Multiple Access (NOMA)  \cite{ LiuHanzoProceedings17, IslamDobreSurvey17} has been proposed for fifth Generation and beyond (B5G) networks to increase  spectral efficiency and serve a larger number of users. The increased users density due to massive use of Internet of Things (IoT) sensors  \cite{LiangNOMA_WCom17,ShirDohlerJSAC17} and their coexistence with enhanced mobile broadband (eMBB) smartphones justify the need for NOMA and for efficient  optimization strategies.  In NOMA, users signals are multiplexed  with superposition coding (SC) at the transmitters and successive interference cancellation (SIC) at the receivers. In this paper, we focus on uplink NOMA, where one Base Station (BS) decodes several users signals by descending order of the received channel gains \cite{HossainIEEEAccess16}. We consider a cell composed of one BS, a  set of IoT sensors and a set of eMBB smartphones. IoT sensors and eMBB users may be multiplexed  with NOMA, depending on whether NOMA will provide a better multi-objective metric  after power optimization than Orthogonal Multiple Access (OMA).  The objective function to be optimized per user is either the spectral efficiency (SE) for eMBB or the energy efficiency (EE) for IoT. Consequently, the multi-objective metric  is the weighted sum of SE and EE.

Few papers in the literature have investigated resource allocation to maximize EE in uplink NOMA. 
In \cite{ZengPoorIoT19},  Dinkelbach algorithm \cite{boyd} is used to   maximize  the global EE subject to minimum data rates per user. In \cite{WangDRLEE20}, three deep reinforcement learning techniques are proposed to maximize the weighted sum of EE subject to a minimum data rate per user.  To the best of our knowledge, multi-objective optimization problems aiming at maximizing both EE and  SE in NOMA have only been studied in the downlink. In \cite{KhanCognitive19,8417647}, multi-objective problems are reformulated as  weighted sum maximization problems, and power allocation is then solved by dual decomposition. 

In this paper, eMBB and IoT users are paired on time slots with the objective to favor NOMA as much as possible.  However, NOMA should only be used if the optimized multi-objective metric after power allocation is larger with NOMA than with OMA. The proposed strategy is therefore an adaptive strategy where NOMA is only used when this is beneficial for the system, as in \cite{PischellaNOMA2019WCL,PischellaCL20}. In order to evaluate when this takes place, we first study all possible multi-objective problems and evaluate when NOMA outperforms OMA. We then deduce a clustering algorithm aiming at maximizing performances while using NOMA as often as possible. The proposed clustering and power optimization algorithm is more efficient than non-adaptive multiple access (MA) techniques and other clustering solutions. We focus on clusters of two users in order to make the best use of NOMA. Larger clusters may indeed lead to error propagation when performing SIC \cite{Liu19_ImperfectSIC} and to additional processing delays for the users whose signal is decoded last. 

This paper is organized as follows: Section \ref{Section_MOP_PA} presents the different multi-objective power allocation problems and their solutions for both NOMA and OMA. Based on these results, the proposed clustering and MA selection algorithm is described in Section \ref{SectionclusteringAlgo}. Its performance results are assessed in Section \ref{Section_NumericalResults} and conclusions are given in Section \ref{Section_Conclusion}.

\section{Multi-objective power allocation problems in NOMA and OMA \label{Section_MOP_PA}}
We consider an uplink two-users systems. The  channel gain to the BS divided by the noise power is denoted as $\gamma_i, i \in \{1,2\}$ and the transmit power as $p_i, i \in \{1,2\}$. We always assume that $\gamma_1 \leq \gamma_2$.  User $1$ is consequently referred to as the weak user, and user $2$ as the strong user. Users $i, i \in \{1,2\}$ may be either IoT or eMBB. If user $i$ is an IoT, it aims at maximizing its energy efficiency, whereas if it is an eMBB, it aims at maximizing its spectral efficiency. The spectral efficiency in bits/s/Hz in NOMA is equal to:
\begin{align}
    SE_{N,2}& = 2\log\left(1+\frac{\gamma_2 p_2}{1+\gamma_1 p_1}  \right);\\
    SE_{N,1}&= 2 \log\left(1+\gamma_1 p_1\right),
\end{align}
where $\log \triangleq \log_2$. The spectral efficiency in OMA is:
\begin{align}
    SE_{O,i}= \log\left(1+  \gamma_i p_i\right) \  \forall i, i \in \{1,2\}. 
\end{align}
The EE in bits/J/Hz in NOMA is defined as:
\begin{align}
    EE_{N,i}= \frac{SE_{N,i}}{2(\phi p_i+Q) } \  \forall i, i \in \{1,2\},
\end{align}
where $\phi$ is the inverse of amplifier efficiency and $Q$ is the circuit power. The EE in OMA is equal to:
\begin{align}
    EE_{O,i}= \frac{SE_{O,i}}{\phi p_i+Q }  \ \forall i, i \in \{1,2\}.
\end{align} 
The multi-objective problem is formulated as a weighted sum objective problem, where weight $w_i$ applies to user  $i, i \in \{1,2\}$, and $w_2 = 1-w_1$.  In this weighted sum objective problem, normalization factors for SE and EE are denoted as $K_S$ and $K_E$, respectively. 
Finally, the maximum total transmit power per user $i \in \{1,2\}$ in NOMA is equal to  $P_i$, whereas the maximum total transmit power per user in OMA is set to $2P_i$. Therefore, the same power budget applies with both MA techniques, since each user in OMA is only active every other time slot. 

In the following, we study the four involved sub-MOP respectively. In each subproblem, both NOMA and OMA are analyzed. 

\subsection{Subproblem $w_1 SE_1+w_2 SE_2$}
This subproblem corresponds to  both users being eMBB.
\subsubsection{NOMA}
The objective function can be expressed as
\begin{align}
\max_{p_1\leq P_1, p_2\leq P_2} \quad &\frac{w_1}{K_S} SE_{N,1} +\frac{w_2}{K_S} SE_{N,2} \\
&=\frac{w_1}{K_S} 2\log\left(1+\gamma_1 p_1\right)\nonumber\\
&+\frac{w_2}{K_S}2\log\left(1+\frac{\gamma_2 p_2}{1+\gamma_1 p_1}\right),\label{sese}
\end{align}
which is not concave. Knowing the fact that $p_2$ should be fixed to be $P_2$, we use the quadratic transform \cite{yuwei} to reformulate \eqref{sese} to a concave function as
\begin{align}\label{rr_noma}
\max_{p_1\leq P_1,y} \quad &\frac{2w_1}{K_S} \log\left(1+\gamma_1 p_1\right)\nonumber\\&+\frac{2w_2}{K_S} \log\left(1+2y\sqrt{\gamma_2 P_2}-y^2\left(1+\gamma_1 p_1\right)\right),\\
s.t. \quad & 2y\sqrt{\gamma_2 P_2}-y^2\left(1+\gamma_1 p_1\right)\geq 0,\label{newcons}
\end{align}
where $y$ is a newly introduced variable. The new constraint \eqref{newcons} is introduced to guarantee the function of the logarithm. Similar with the proof in \cite{yuwei} and \cite{wcnc2020}, it can be proven that the reformulated problem \eqref{rr_noma} is concave.

As in \cite{yuwei} and \cite{wcnc2020}, variables $p_1$ and $y$ are iteratively updated to reach a stationary point of the original problem in \eqref{sese}, where $p_1$ is optimized by Karush–Kuhn–Tucker (KKT) conditions and $y$ is updated by $y=\frac{\sqrt{\gamma_2 P_2}}{1+\gamma_1 p_1}$.

As we will see in the following, similar methods are used to solve the other subproblems for NOMA in this paper. 

\subsubsection{OMA}
\begin{align}
\max_{p_1\leq 2P_1, p_2\leq 2P_2} &\frac{w_1}{K_S} SE_{O,1} +\frac{w_2}{K_E} SE_{0,2} \\
&=\frac{w_1}{K_S} \log\left(1+\gamma_1 p_1\right)+\frac{w_2}{K_E}\log\left(1+\gamma_2 p_2\right),
\end{align}
which obviously equals to
\begin{equation}
\frac{w_1}{K_S} \log\left(1+\gamma_1\cdot 2P_1\right)+\frac{w_2}{K_E}\log\left(1+\gamma_2\cdot 2P_2\right).
\end{equation}

\subsection{Subproblem $w_1 EE_1+w_2 SE_2$}
This subproblem is considered if the cluster consists of an eMBB and an IoT, and the eMBB  has the highest SNR.
\subsubsection{NOMA}
The objective function can be expressed as 
\begin{align}
\max_{p_1\leq P_1, p_2\leq P_2} \quad &\frac{w_1}{K_E} EE_{N,1} +\frac{w_2}{K_S} SE_{N,2} \\
&=\frac{w_1}{K_E} \frac{\log\left(1+\gamma_1 p_1\right)}{\phi p_1+Q}\nonumber\\
&+\frac{w_2}{K_S}2\log\left(1+\frac{\gamma_2 p_2}{1+\gamma_1 p_1}\right),
\end{align}
which is, similar with the transform from \eqref{sese} to \eqref{rr_noma}, equivalent with 
\begin{align}
\max_{p_1\leq P_1,\bm{y}} \quad
&\frac{w_1}{K_E} 2 y_1\sqrt{\log\left(1+\gamma_1 p_1\right)}-\frac{w_1}{K_E}y_1^2\left(\phi p_1+Q\right)\nonumber\\&+\frac{w_2}{K_S}2\log\left(1+2y_2\sqrt{\gamma_2 P_2}-y_2^2\left(1+\gamma_1 p_1\right)\right),\\
s.t. \quad & 2y_2\sqrt{\gamma_2 P_2}-y_2^2\left(1+\gamma_1 p_1\right)\geq 0,
\end{align}
where $\bm{y}=[y_1,y_2]$ is an introduced  auxiliary variable vector. $p_1$ can be optimized by KKT conditions and the updates are respectively $y_1=\frac{\sqrt{\log\left(1+\gamma_1 p_1\right)}}{\phi p_1+Q}$ and $y_2=\frac{\sqrt{\gamma_2 P_2}}{1+\gamma_1 p_1}$.

\subsubsection{OMA}
\begin{align}
\max_{p_1\leq 2P_1, p_2\leq 2P_2} \quad &\frac{w_1}{K_E} EE_{O,1} +\frac{w_2}{K_S} SE_{O,2} \\
&=\frac{w_1}{K_E} \frac{\log\left(1+\gamma_1 p_1\right)}{\phi p_1+Q}+\frac{w_2}{K_S}\log\left(1+\gamma_2 p_2\right),
\end{align}
which can be rewritten as
\begin{align}
\max_{p_1\leq 2P_1} \frac{w_1}{K_E} \frac{\log\left(1+\gamma_1 p_1\right)}{\phi p_1+Q}+\frac{w_2}{K_S}\log\left(1+\gamma_2 2P_2\right).
\end{align}
Only the first term needs to be optimized, which is as simple as a Dinkelbach's algorithm \cite{boyd} such that 
\begin{align}
\max_{p_1\leq 2P_1} \log\left(1+\gamma_1 p_1\right)-\lambda \left(\phi p_1+Q\right),
\end{align}
where $\lambda$ is iteratively updated to maximize the EE of user 1.

\subsection{Subproblem $w_1 SE_1+w_2 EE_2$}
This subproblem models clusters with an eMBB and an IoT user, where the IoT  has the largest SNR.
\subsubsection{NOMA}
The objective function is equal to
\begin{align}
\max_{p_1\leq P_1,p_2\leq P_2} & \quad \frac{w_1}{K_S} SE_{N,1} +\frac{w_2}{K_E} EE_{N,2} \\
&=\frac{w_1}{K_S} 2\log\left(1+\gamma_1 p_1\right)+\frac{w_2}{K_E}\frac{\log\left(1+\frac{\gamma_2 p_2}{1+\gamma_1 p_1}\right)}{\phi p_2+Q},
\end{align}
which is, similar with the transform from \eqref{sese} to \eqref{rr_noma}, equivalent with
\begin{align}
\max_{p_1\leq P_1,p_2\leq P_2,y,t} & \quad  \frac{w_1}{K_S} 2\log\left(1+\gamma_1 p_1\right)\nonumber\\
&+\frac{w_2}{K_E}2y\sqrt{\log\left(1+2t\sqrt{\gamma_2 p_2}-t^2\left(1+\gamma_1 p_1\right)\right)}\nonumber\\
&-\frac{w_2}{K_E}y^2\left(\phi p_2+Q\right)\\
s.t.\quad & 2t\sqrt{\gamma_2 p_2}-t^2\left(1+\gamma_1 p_1\right)\geq 0,
\end{align}
where $(y,t)$ are newly introduced variables. $p_1$ and $p_2$ can be optimized by KKT conditions. The auxiliary variables are then updated respectively as $t=\frac{\sqrt{\gamma_2 p_2}}{1+\gamma_1 p_1}$ and $y=\frac{\sqrt{\log\left(1+2t\sqrt{\gamma_2 p_2}-t^2\left(1+\gamma_1 p_1\right)\right)}}{\phi p_2+Q}$.

\subsubsection{OMA}
\begin{align}
\max_{p_1\leq 2P_1, p_2\leq 2P_2} & \quad \frac{w_1}{K_S} SE_{O,1} +\frac{w_2}{K_E} EE_{O,2} \\
&=\frac{w_1}{K_S} \log\left(1+\gamma_1 p_1\right)+\frac{w_2}{K_E}\frac{\log\left(1+\gamma_2 p_2\right)}{\phi p_2+Q},
\end{align}
which is equivalent with 
\begin{equation}
\max_{p_2\leq 2P_2} \frac{w_1}{K_S} \log\left(1+\gamma_1 2P_1\right)+\frac{w_2}{K_E}\frac{\log\left(1+\gamma_2 p_2\right)}{\phi p_2+Q}.
\end{equation}
Only the second term needs to be optimized, which is as simple as 
\begin{align}
\max_{p_2\leq 2P_2} \quad \log\left(1+\gamma_2 p_2\right)-\lambda \left(\phi p_2+Q\right).
\end{align}

\subsection{Subproblem $w_1 EE_1+w_2 EE_2$}
This final case corresponds to a cluster of two IoT.
\subsubsection{NOMA}
The objective function for NOMA is expressed as
\begin{align}
\max_{p_1\leq P_1,p_2\leq P_2} & \quad \frac{w_1}{K_E} EE_{N,1} +\frac{w_2}{K_E} EE_{N,2} \\
&=\frac{w_1}{K_E} \frac{\log\left(1+\gamma_1 p_1\right)}{\phi p_1+Q}+\frac{w_2}{K_E}\frac{\log\left(1+\frac{\gamma_2 p_2}{1+\gamma_1 p_1}\right)}{\phi p_2+Q}.\label{ee_ee_noma}
\end{align}
\subsubsection{OMA}
The objective function for OMA is expressed as
\begin{align}
\max_{p_1\leq 2P_1,p_2\leq 2P_2} & \quad \frac{w_1}{K_E} EE_{O,1} +\frac{w_2}{K_E} EE_{O,2} \\
&=\frac{w_1}{K_E} \frac{\log\left(1+\gamma_1 p_1\right)}{\phi p_1+Q}+\frac{w_2}{K_E}\frac{\log\left(1+\gamma_2 p_2\right)}{\phi p_2+Q}.\label{ee_ee_oma}
\end{align}
First, with the same value of $p_1$ and $p_2$, \eqref{ee_ee_noma} is always smaller than \eqref{ee_ee_oma}. In addition, \eqref{ee_ee_noma} is a subset of \eqref{ee_ee_oma}. Therefore, we can conclude that OMA is always better than NOMA when both users are IoT.

\section{Proposed clustering algorithm \label{SectionclusteringAlgo}}
In this section, we propose a clustering algorithm that selects  users so that NOMA is  optimal for as many pairs as possible after power optimization. For that purpose, we first compute the obtained metric after power optimization for the four multi-objective problems detailed in Section \ref{Section_MOP_PA}. 
\subsection{Best MA strategy depending on the MOP}
\begin{figure}[h]
	\centering
	\includegraphics[width=3.5in]{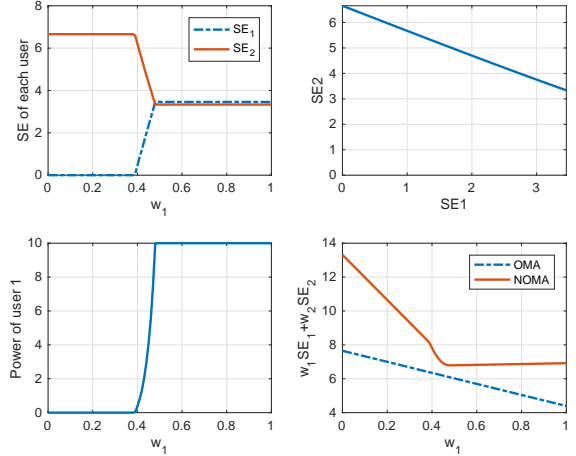}
	\caption{Subproblem $w_1 SE_1+w_2 SE_2$. Subfigures are respectively: individual SE for NOMA, SE of user 2 vs. user 1 for NOMA, transmit power of user 1 for NOMA (user 2 always uses full power), and the comparison between NOMA and OMA. For NOMA, user 1 is first switched off and then switched on if $w_1$ is larger than a threshold.}
	\label{f1}
\end{figure}

\begin{figure}
	\centering
	\includegraphics[width=3.5in]{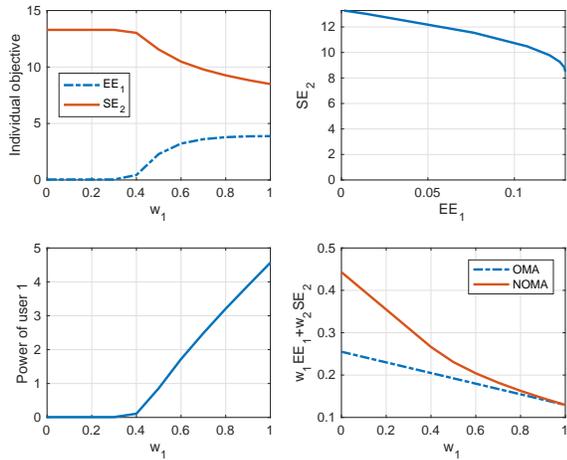}
	\caption{Subproblem $w_1 EE_1+w_2 SE_2$. Subfigures are respectively: EE of user 1 and SE of user 2 for NOMA, SE of user 2 vs. EE of user 1 for NOMA, transmit power of user 1 for NOMA (user 2 always uses full power), and the comparison between NOMA and OMA. For NOMA, user 1 is first inactive and then gradually increases its power. The two MA converge to the same performance when $w_1=1$ because $p_2=0$ at this point.}
	\label{f2}
\end{figure}

\begin{figure}[t]
	\centering
	\includegraphics[width=3.5in]{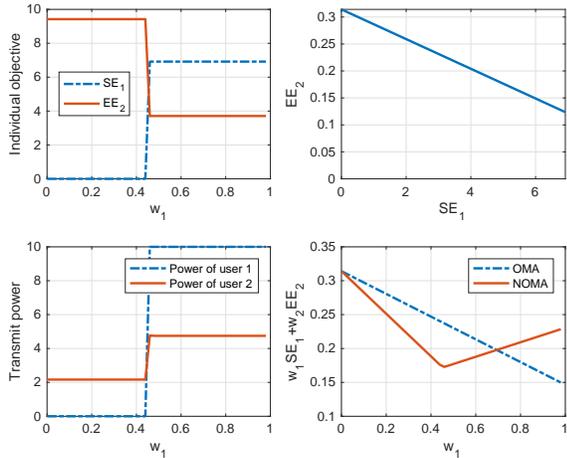}
	\caption{Subproblem $w_1 SE_1+w_2 EE_2$. Subfigures are respectively: SE of user 1 and EE of user 2 for NOMA, EE of user 2 vs. SE of user 1 for NOMA, transmit power for NOMA, and the comparison between NOMA and OMA. For NOMA, user 1 is first switched off and then switched on if $w_1$ is larger than a threshold, while user 2 increases a bit its power.}
	\label{f3}
\end{figure}

In order to evaluate the influence of MA strategy and of clustering if NOMA is used, we first consider two users where user 2 has a channel gain $10$dB higher than that of user 1. The power allocation results for all MOP except the one where both users maximize their EE are represented on Fig. \ref{f1},  Fig. \ref{f2}, and Fig. \ref{f3}. If the MOP is the weighted sum of EE, as already seen, OMA always outperforms NOMA.

\begin{figure}
	\centering
	\includegraphics[width=3.5in]{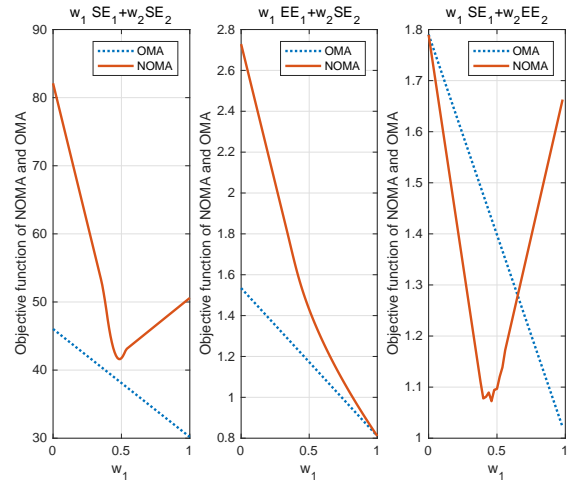}
	\caption{Comparison between NOMA and OMA of the three subproblems with 10 users.}
	\label{combined}
\end{figure}

Then we consider $10$ users within the cell, such that distances between user $k$ and the BS is $d_k=(10(11-k))$m  and the path loss exponent is $n=2$. The normalized channel of user $k$ is assumed to be $10000/d_k^2$ and no fading is assumed.  In NOMA, users are clustered so as to maximize the signal to interference plus noise ratio of the strong user: $k = {1,...,5}$, user $k$ is matched with user $ 10-k$. The number of time slots is equal to $10$ and if NOMA is used, each pair occupies $2$ consecutive time slots. The power allocation results are shown in Fig. \ref{combined}, which is consistent with Fig. \ref{f1}, \ref{f2}, and \ref{f3}.

The best MA strategy depending on the MOP is summarized in Table \ref{TableBestStrategy}.
 \begin{table}[h]
\renewcommand{\arraystretch}{1.3}
\caption{Best MA strategy depending on the subproblem}
\label{TableBestStrategy}
\begin{center}
\begin{tabular}{ |c|c||c|}
 \hline
 Weak user & Strong user & Best MA strategy  \\
      \hline  \hline 
eMBB &    eMBB  & NOMA \\
    \hline
IoT  & eMBB   & NOMA \\
     \hline
eMBB & IoT & NOMA when  $w_1 \geq w_{1,\text{min}}$, OMA otherwise \\
         \hline
   IoT  & IoT   & OMA \\    
    \hline
 \end{tabular}
\end{center}
\end{table}

\subsection{Clustering algorithm to maximize NOMA use}
The clustering algorithm based on the results from Table \ref{TableBestStrategy} is detailed hereafter. 

Let us assume that the cell contains  $N_{IoT}$ IoT   and $N_{eMBB}$ eMBB. Let $g_m$ be the channel gain between the $m^{\text{th}}$ IoT  and the BS and $h_m$  be the channel gain between the $m^{\text{th}}$ eMBB  and the BS. Users are ordered by descending channel gains: $g_1 \geq g_2 \geq ... \geq g_{N_{IoT}}$ and $h_1 \geq h_2 \geq ... \geq h_{N_{eMBB}}$. 
Let us define  $M = \min\{N_{IoT},N_{eMBB}\}$ and  $l^*$ as  follows:
\begin{align}
l^*= \text{arg max}_{1 \leq l \leq M} \ g_{(N_{IoT}+1-l)}<h_l
\end{align}
Then the following inequalities stand: 
$ h_1 \geq h_2 \geq ... \geq h_l^* \geq g_{(N_{IoT}+1-l^*)} \geq g_{(N_{IoT}+2-l^*)} \geq... \geq g_{N_{IoT}}$
and:
$ g_1 \geq g_2 \geq ... \geq g_{(N_{IoT}-l^*)} \geq h_{l^*+1} \geq h_{l^*+2}  \geq ... \geq h_{N_{eMBB}}$.

 The first set of clustered users is composed of weak IoT users and strong eMBB users and contains the following pairs:
\begin{align}\label{cluster1}
 \mathcal{S}_1 = \{(h_1, g_{N_{IoT}}); (h_2, g_{(N_{IoT}-1)});...; 
 (h_{l^*}, g_{(N_{IoT}+1-l^*)})\}
\end{align}
The pairs in  $\mathcal{S}_1$ aim at maximizing $w_1 EE_1 + w_2 SE_2$ with NOMA, as this is the best strategy for this MOP according to Table \ref{TableBestStrategy} .

 The second set of clustered users is composed of weak eMBB users and strong IoT users paired as follows: 
\begin{align}\label{cluster2}
 \mathcal{S}_2 = & \left\{( h_{N_{eMBB}}, g_{1}), (h_{N_{eMBB}-1}; g_{2});...;  \right. \nonumber \\ 
&\left. (h_{(N_{eMBB} -M + l^*+1)},  g_{(M-l^*)})\right\}
\end{align}
The pairs in  $\mathcal{S}_2$  aim at maximizing $w_1 SE_1 + w_2 EE_2$ with NOMA if $w_1 \geq w_{1,\text{min}}$ and with OMA otherwise.

Finally, if $M= N_{IoT}$, the  eMBB that do not belong to $\mathcal{S}_1 \cup  \mathcal{S}_2$ are paired by descending index and the objective function is  $w_1 SE_1 + w_2 SE_2$ with NOMA for all pairs. If the number of elements is odd, then one eMBB user is in OMA.

Similarly, if $M= N_{eMBB}$, the IoT users that do not belong to $\mathcal{S}_1 \cup  \mathcal{S}_2$   are   paired and their objective function is  $w_1 EE_1 + w_2 EE_2$ with OMA.  OMA is also used if the number of elements is odd.

\section{Numerical results \label{Section_NumericalResults}}
In this section, we compare the proposed clustering, MA strategy selection and power allocation algorithm with two other algorithms:
\begin{itemize}
\item In the first algorithm, we assume that the proposed clustering algorithm is used, but that all clusters transmit with the same MA strategy.
\item In the second algorithm, we assume that clustering is randomly performed, and followed by either OMA or NOMA for all clusters.  
\end{itemize} 

\begin{figure}[h]
	\centering
	\includegraphics[width=3.5in]{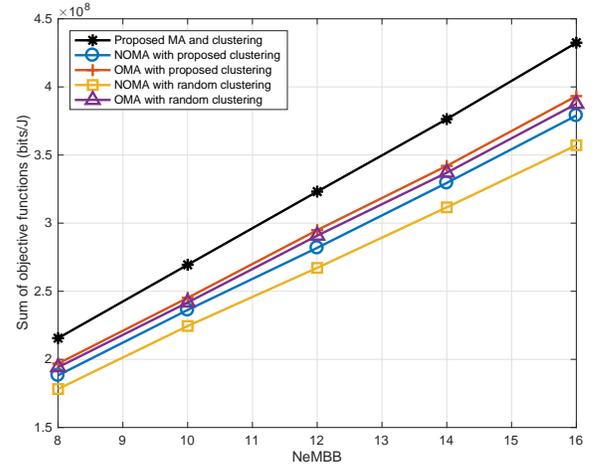}
	\caption{Performance vs. $N_{eMBB}=N_{IoT}$ when $w_1=0.4$ and $K_S=30, K_E=1$.}
	\label{f9}
\end{figure}

\begin{figure}[h]
	\centering
	\includegraphics[width=3.5in]{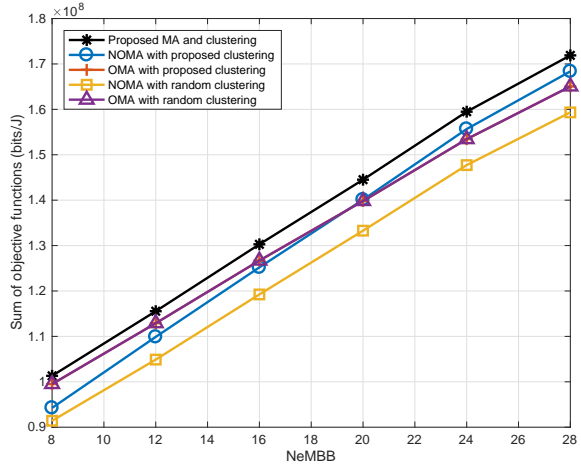}
	\caption{Performance vs. $N_{eMBB}$ when  $N_{IoT}=6$ and $w_1=0.5$. $K_S=100$ and $K_E=1$.}
	\label{f10}
\end{figure}

\begin{figure}[h]
	\centering
	\includegraphics[width=3.5in]{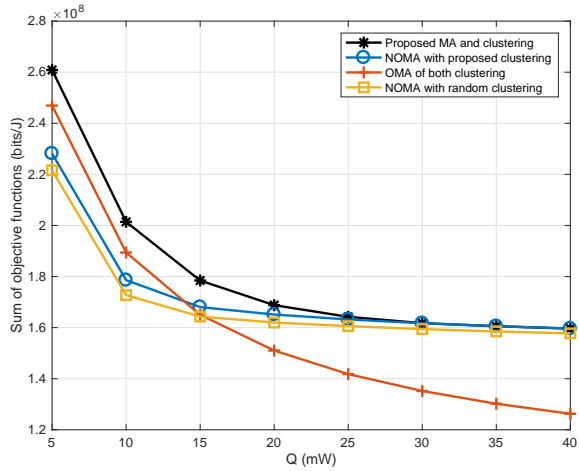}
	\caption{Performance vs. $Q$. $N_{IoT}=8$, $N_{eMBB}=8$ and $w_1=0.5$. $K_S=30$ and $K_E=1$.}
	\label{f11}
\end{figure}

\begin{figure}[h]
	\centering
	\includegraphics[width=3.5in]{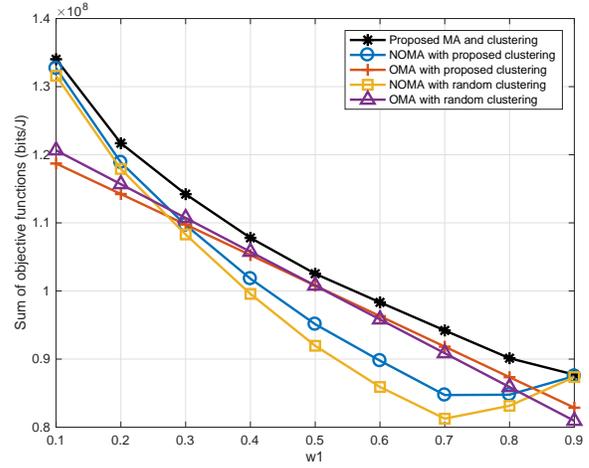}
	\caption{Performance vs. $w_1$. $N_{IoT}=6$ and $N_{eMBB} = 8$.  $K_S=100$ and $K_E=1$.}
	\label{f12}
\end{figure}

All users are randomly located within the cell whose outer radius is equal to $100$m and inner radius is equal to $10$m.  We consider the following values for the channel modeling of path loss: $-G_0+10n\log_{10}(d)$, where $n=2$,  $G_0=-(G_1+M_l)=-70$dB, where $G_1 = 30$dB is the gain factor at d = 1m and $M_l = 40$dB \cite{wcnc2020}. The receiving noise power is $B\cdot(N_0+N_f)$, where the noise power spectral density is set to $N_0=-170$ dBm/Hz,  and the noise figure to $N_f = 10$dB/Hz. The bandwidth per user is equal to $B=100$ kHz. On Fig. \ref{f9} to \ref{f12}, SE and EE are given by taking into account this bandwidth and are respectively expressed in bits/s and bits/J.

We assume that the inverse of amplifier efficiency is equal to $\phi=2$, the circuit power is $Q=10$ mW, and the transmit power budget of each user is $P_1=P_2=10$ mW. We set the normalization factors for SE and EE ($K_S$ and $K_E$) individually in each figure. $K_S$ is used to make the spectral efficiency value of the same magnitude as that of the energy efficiency. Therefore, $K_S$ can be viewed as a fixed power consumption. A larger value of $K_S$ means the IoT users have higher priority. Consequently, the units of the objective functions in the following figures are bits/J.

In Fig. \ref{f9}, we plot the sum of objective functions versus the number of users in each group (eMBB users group and IoT users group). We assume that $N_{eMBB} = N_{IoT}$. Thus, all users belong to  $\mathcal{S}_1 \cup \mathcal{S}_2$. The figure shows that our proposed clustering and MA strategy always outperforms the others. From the former numerical analysis in Fig. \ref{combined}, the fact that OMA outperforms NOMA implies that the advantage of OMA in the clusters where the MOP objective function is $w_1 SE_1+w_2 EE_2$ is more dominating.

In Fig. \ref{f10}, contrary to Fig. \ref{f9}, we fix  $N_{IoT}= 6$ and increase the number of eMBB users, $N_{eMBB}$. Unlike in Fig. \ref{f9} where  the proposed clustering  followed by OMA is always better than NOMA, it is observed in Fig. \ref{f10} that the proposed clustering followed by NOMA outperforms OMA when  $N_{eMBB}$  is large. In this case indeed, the  eMBB users that do not belong to $\mathcal{S}_1 \cup \mathcal{S}_2$ form the clusters for which  subproblem 1 ($w_1 SE_1+w_2 SE_2$) is solved. As seen in Table \ref{TableBestStrategy},  NOMA has a better achievable SE than OMA for this MOP. The two curves of OMA have exactly the same values because $w_1=0.5$ keeps the objective functions of all users the same for the two MA clustering, which will be further confirmed by Fig. \ref{f12}.

In Fig. \ref{f11}, we change the circuit power of each user. While the proposed clustering always has a better performance, we see that the proposed scheme converges to NOMA of both clustering, because when the objective function of eMBB users are quite dominating for large circuit power, clustering does not play an important role and NOMA has an advantage than OMA for eMBB users.

Finally in Fig. \ref{f12}, performances of various clustering strategies and MA schemes are compared with respect to the weight $w_1$. It is observed that the proposed clustering strategy has a better performance for all values of $w_1$, more specifically for small and medium values of $w_1$. Please note that small $w_1$ implies focusing on optimizing the strongest channel. This is consistent with our motivation to propose the clustering method, which is to pair the strongest channel with the weakest channel to suppress interference so that the users with strong channels reach higher performance. For large $w_1$ values, the random clustering performs similar to the proposed one, because more strong channels are optimized since some users with higher gain might be the weakest user (user 1) in a cluster for random clustering, which can never happen in the proposed clustering. By comparing the proposed clustering followed by NOMA (blue curve) with the proposed clustering followed by OMA (red curve), we observe that NOMA outperforms OMA for small and large values of $w_1$, which is quite consistent with the numerical analysis in Fig. \ref{combined}. 

\section{Conclusion \label{Section_Conclusion}}
This paper has proposed an adaptive multiple access strategy combined with clustering and power optimization for multi-objective optimization problems that model a combination of IoT and eMBB users coexisting within a cell.  We showed through extensive simulations that the proposed strategy outperforms non adaptive MA and other clustering algorithms. In the context of dense networks with massive numbers of IoT and heterogeneous users' Quality of Service, such joint MA, clustering and power optimization strategies can consequently provide large benefits. 

\section*{Acknowledgment}
This work was conducted while Mylene Pischella was a visiting researcher at Universit{\'e} catholique de Louvain and at ETIS UMR8051, CY University, ENSEA, CNRS, F-95000, Cergy, France.

This work was supported by FNRS (Fonds National de la recherche scientifique) under EOS project Number 30452698. The authors would like to thank UCLouvain for funding the ARC SWIPT project.
\bibliographystyle{IEEEbib}

\bibliography{BIBLIO_NOMA_Adaptive.bib}

\begin{thebibliography}{10}

\bibitem{LiuHanzoProceedings17}
Y.~{Liu}, Z.~{Qin}, M.~{Elkashlan}, Z.~{Ding}, A.~{Nallanathan}, and
  L.~{Hanzo},
\newblock ``{Nonorthogonal Multiple Access for 5G and Beyond},''
\newblock {\em Proceedings of the IEEE}, vol. 105, no. 12, pp. 2347--2381, Dec
  2017.

\bibitem{IslamDobreSurvey17}
S.~M.~R. {Islam}, N.~{Avazov}, O.~A. {Dobre}, and K.~{Kwak},
\newblock ``{Power-Domain Non-Orthogonal Multiple Access (NOMA) in 5G Systems:
  Potentials and Challenges},''
\newblock {\em IEEE Communications Surveys Tutorials}, vol. 19, no. 2, pp.
  721--742, Secondquarter 2017.

\bibitem{LiangNOMA_WCom17}
Y.~{Liang}, X.~{Li}, J.~{Zhang}, and Z.~{Ding},
\newblock ``{Non-Orthogonal Random Access for 5G Networks},''
\newblock {\em IEEE Transactions on Wireless Communications}, vol. 16, no. 7,
  pp. 4817--4831, July 2017.

\bibitem{ShirDohlerJSAC17}
M.~{Shirvanimoghaddam}, M.~{Condoluci}, M.~{Dohler}, and S.~J. {Johnson},
\newblock ``{On the Fundamental Limits of Random Non-Orthogonal Multiple Access
  in Cellular Massive IoT},''
\newblock {\em IEEE Journal on Selected Areas in Communications}, vol. 35, no.
  10, pp. 2238--2252, Oct 2017.

\bibitem{HossainIEEEAccess16}
M.~S. Ali, H.~Tabassum, and E.~Hossain,
\newblock ``Dynamic user clustering and power allocation for uplink and
  downlink non-orthogonal multiple access (noma) systems,''
\newblock {\em IEEE Access}, vol. 4, pp. 6325--6343, 2016.

\bibitem{ZengPoorIoT19}
M.~{Zeng}, A.~{Yadav}, O.~A. {Dobre}, and H.~V. {Poor},
\newblock ``Energy-efficient joint user-rb association and power allocation for
  uplink hybrid noma-oma,''
\newblock {\em IEEE Internet of Things Journal}, vol. 6, no. 3, pp. 5119--5131,
  2019.

\bibitem{boyd}
S.~Boyd and L.~Vandenberghe,
\newblock ``Convex optimization,''
\newblock {\em Cambridge University Press}, 2004.

\bibitem{WangDRLEE20}
X.~{Wang}, Y.~{Zhang}, R.~{Shen}, Y.~{Xu}, and F.~{Zheng},
\newblock ``Drl-based energy-efficient resource allocation frameworks for
  uplink noma systems,''
\newblock {\em IEEE Internet of Things Journal}, pp. 1--1, 2020.

\bibitem{KhanCognitive19}
W.~U. {Khan}, F.~{Jameel}, T.~{Ristaniemi}, S.~{Khan}, G.~A.~S. {Sidhu}, and
  J.~{Liu},
\newblock ``Joint spectral and energy efficiency optimization for downlink noma
  networks,''
\newblock {\em IEEE Transactions on Cognitive Communications and Networking},
  pp. 1--1, 2019.

\bibitem{8417647}
D.~{Ni}, L.~{Hao}, X.~{Qian}, and Q.~T. {Tran},
\newblock ``Energy-spectral efficiency tradeoff of downlink noma system with
  fairness consideration,''
\newblock in {\em 2018 IEEE 87th Vehicular Technology Conference (VTC Spring)},
  2018, pp. 1--5.

\bibitem{PischellaNOMA2019WCL}
M.~{Pischella} and D.~{Le Ruyet},
\newblock ``{NOMA-Relevant Clustering and Resource Allocation for Proportional
  Fair Uplink Communications},''
\newblock {\em IEEE Wireless Commun. Lett.}, vol. 8, no. 3, pp. 873--876, June
  2019.

\bibitem{PischellaCL20}
M.~{Pischella}, A.~{Chorti}, and I.~{Fijalkow},
\newblock ``{Performance Analysis of Uplink NOMA-Relevant Strategy Under
  Statistical Delay QoS Constraints},''
\newblock {\em IEEE Wireless Communications Letters}, pp. 1--1, 2020.

\bibitem{Liu19_ImperfectSIC}
M.~{Liu}, T.~{Song}, and G.~{Gui},
\newblock ``Deep cognitive perspective: Resource allocation for noma-based
  heterogeneous iot with imperfect sic,''
\newblock {\em IEEE Internet of Things Journal}, vol. 6, no. 2, pp. 2885--2894,
  2019.

\bibitem{yuwei}
K.~Shen and W.~Yu,
\newblock ``Fractional programming for communication systems—part i: Power
  control and beamforming,''
\newblock {\em IEEE Trans. Signal Process.}, vol. 66, no. 10, pp. 2616--2630,
  May 2018.

\bibitem{wcnc2020}
Z.~Wang, L.~Vandendorpe, M.~Ashraf, Y.~Mou, and N.~Janatian,
\newblock ``Minimization of sum inverse energy efficiency for multiple base
  station systems,''
\newblock in {\em 2020 IEEE Wireless Communications and Networking Conference
  (WCNC)}, 2020.

\end{thebibliography}



\end{document}